\documentclass[reprint,prb,aps,twocolumn,superscriptaddress,10pt]{revtex4-2}
\usepackage{amsmath}
\usepackage{graphicx}
\usepackage{color}
\usepackage{sidecap}
\usepackage[dvipsnames]{xcolor}

\usepackage[colorlinks,
            citecolor = blue, 
            linkcolor = blue,
            urlcolor  = blue,
            anchorcolor = blue,
            breaklinks = true
            ]{hyperref}

\usepackage{cleveref}
\usepackage{soul}
\usepackage{ amssymb }

\newif\ifdraft
\drafttrue

\begin{document}


\title{Nonequilibrium dynamics of doped Chern ferromagnets: a case study for false vacuum decay}

\author{Kilian Kuhlbrodt}
\affiliation{Institute for Quantum Electronics, ETH Zürich, Zürich, Switzerland}
\author{Olivier Huber}
\affiliation{Institute for Quantum Electronics, ETH Zürich, Zürich, Switzerland}
\author{Oliver Breach}
\affiliation{Rudolf Peierls Centre for Theoretical Physics, Clarendon Laboratory, Parks Road, Oxford, OX1 3PU, UK}
\author{Weijie Li}
\affiliation{Department of Physics, University of Washington, Seattle, WA, USA}
\author{Xiaodong Xu}
\affiliation{Department of Physics, University of Washington, Seattle, WA, USA}
\author{Kenji Watanabe}
\affiliation{Research Center for Electronic and Optical Materials, National Institute for Materials Science, 1-1 Namiki, Tsukuba 305-0044, Japan}
\author{Takashi Taniguchi}
\affiliation{Research Center for Materials Nanoarchitectonics, National Institute for Materials Science,  1-1 Namiki, Tsukuba 305-0044, Japan}
\author{Martin~Kroner}
\affiliation{Institute for Quantum Electronics, ETH Zürich, Zürich, Switzerland}
\author{Siddharth A. Parameswaran}
\affiliation{Rudolf Peierls Centre for Theoretical Physics, Clarendon Laboratory, Parks Road, Oxford, OX1 3PU, UK}
\author{Atac {\.I}mamo{\u{g}}lu}
\affiliation{Institute for Quantum Electronics, ETH Zürich, Zürich, Switzerland}

\maketitle


{\bf 
Even though metastable false vacuum decay is ubiquitous in physics, its underlying dynamics are still not well understood. Dissipative state preparation in moiré quantum materials provides an exceptional setting for exploring this physics since it allows the possibility of generating exotic quantum states that are not the ground state of the system Hamiltonian. Motivated by recent experiments demonstrating steady-state optical orientation of the spin-valley degree of freedom of holes, here we investigate dynamics of itinerant and Chern ferromagnets in the presence of an opposing magnetic field. Optical pumping using a circularly polarized Laguerre-Gauss beam allows us to deterministically prepare a true vacuum  {\sl bubble} embedded inside a metastable state. Depending on its initial size controlled by the pump power, we observe that the bubble collapses or expands due to an interplay between domain wall and bulk dynamics. For external magnetic fields comparable to the coercive field of ferromagnetism, we observe up to two-orders-of-magnitude prolongation of the spin polarization decay time at commensurate fillings corresponding to integer and fractional Chern insulator states. Our experiments reveal that the nonequilibrium dynamics of the ferromagnetic domains is substantially more sensitive to the precise filling factor around Chern insulator states than standard transport or optical measurements.
} \\

{\bf Introduction}

The decay of a metastable state --- evocatively termed ``the fate of the false vacuum'' in its quantum field theoretic setting --- is a longstanding problem in physics, with applications to fields ranging from cosmology to classical magnetism. As an intrinsically non-equilibrium problem controlled by the interplay of the nucleation --- quantum or thermal --- and the subsequent motion and eventual coalescence of bubbles of the true vacuum, it is notoriously difficult to model in full generality. This challenge has motivated theoretical approaches of increasing sophistication, ranging from Kramers' classic description of diffusive escape over a barrier \cite{Kramers1940} to Langer’s theory of nucleation and metastable-state decay in many-body systems \cite{Langer1969}. The recent advent of quantum simulators capable of realizing false vacuum decay in idealized settings with coherent unitary quantum dynamics has led to a renaissance in its study \cite{Song2022DiscontinuousQPTDrivenOpticalLattice,Zhu2024ColdAtomGaugeFalseVacuum, Cominotti2025TemperatureEffectsFalseVacuumDecay,Luo2025BubbleNucleationQuantumPhaseTransition,Vodeb2025,Osterholz2026CollectiveClusterNucleation,Chao2026RydbergFalseVacuum,Sivasankar2026,Humar2026}. However, probing false vacuum decay experimentally remains challenging in its arguably most prevalent setting --- namely, solid-state systems with realistic dissipation and disorder ---  particularly in a manner that allows spatially-resolved investigation of the post-nucleation dynamics of metastable domains.

Optical control of moiré ferromagnetism provides a powerful new route to address these questions. In twisted molybdenum ditelluride (t-MoTe$_2$) and other semiconductor moiré materials, narrow electronic bands combined with strong interactions stabilize spin-valley polarized Stoner-like ferromagnetic states that spontaneously break time-reversal symmetry. The resulting orbital Ising ferromagnetism occurs in bands with nonzero Chern number, giving rise to chiral transport at sample boundaries and interfaces between oppositely polarized Ising domains --- features that are particularly sharp in integer or fractional quantum anomalous Hall states~\cite{Neupert_PRL_2011,Bernevig_PRX_2011,Cai_Xu_2023, Park_Xu_2023,Zeng_Nature_2023,Xu_APS_2023}. Crucially for our current purposes, recent experiments have demonstrated switching of the Chern number of these fractional (FCI) and integer (ICI) Chern insulator states in a diffraction limited optical spot using optical excitation of an excitonic resonance~\cite{Huber_Nature_2026,Holtzmann_Nature_2026,cai_optical_2026}.

Here, we report time-resolved optical measurements where we use Laguerre-Gauss (LG) laser pulses with a finite orbital angular momentum to write sub-wavelength ferromagnetic domains whose size is controlled by the laser pulse energy and whose spin orientation is opposite to that of the surrounding region. Simultaneously, an external magnetic field is used to tune the energy landscape to select the true global minimum and to ensure that the optically generated sub-wavelength domain ({\sl bubble}) constitutes a region of true vacuum, surrounded by an extensive region of false vacuum. By controlling the size of the optically written domain through the energy of the LG laser pulse and contrasting the resulting magnetization dynamics as we dope away from commensurate filling factors corresponding to ICI or FCI states and across a range of applied fields, we uncover two distinct dynamical regimes of the  low-temperature metastable Ising domains: Near the coercive field, where spontaneous nucleation and domain wall depinning determine the dynamics, the false vacuum decay is dramatically slowed down at the commensurate fillings of the ICI and FCI states ($\nu=-1$ and $\nu=-2/3$) compared to the adjacent states with incommensurate fillings.
In contrast, for magnetic fields well below and well above the coercive field --- where the dynamics is determined by the kinetics of the domain wall --- we observe a much weaker distinction at and away from commensurate fillings. Our work establishes t-MoTe$_2$ and related moiré materials as platforms for investigating phase-conversion dynamics at the intersection of correlations and topology, while enabling sub-micrometer control of topological states. \\

{\bf Optical spin orientation in t-MoTe$_2$}

\begin{figure*}[t]
	\includegraphics[width=0.99\textwidth]{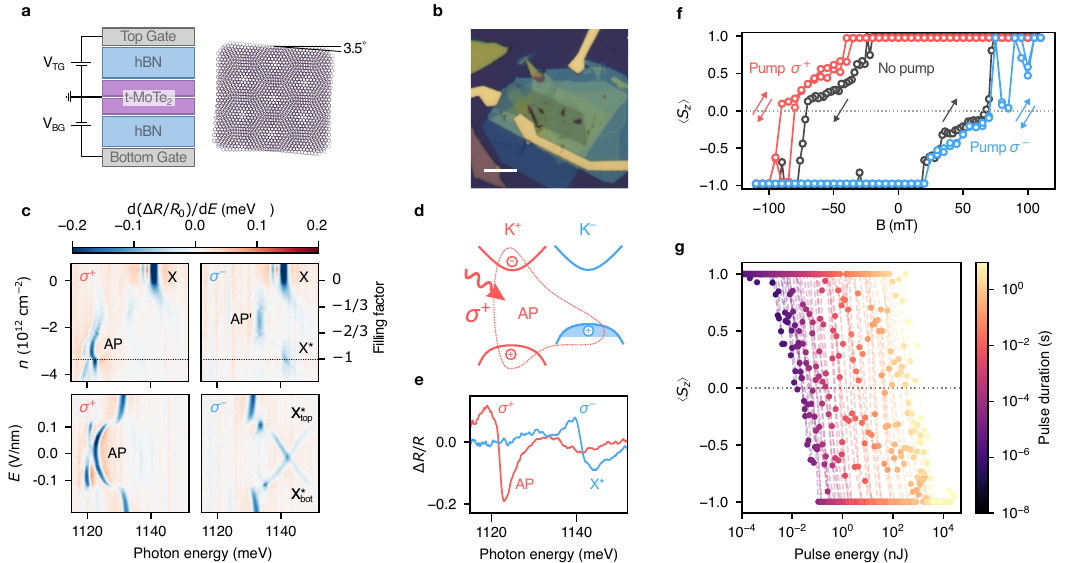}
	\caption{{\bf Optical spin orientation in t-MoTe$_2$ bilayers.} 
    ({\bf a}) Schematic of the device and underlying moiré superlattice. The heterostructure comprises a 3.5$^\circ$-twisted MoTe$_2$ homobilayer encapsulated in hexagonal boron-nitride (hBN) with transparent few-layer graphene gates. 
    ({\bf b}) Optical micrograph of the device. 
    ({\bf c}) Differentiated reflectance contrast spectra $\mathrm{d}(\Delta R/R_0)/\mathrm{d}E$ resolved by circular polarization as a function of $\nu$ at fixed $D \approx 0$ (top panel) and a function of $D$ at fixed $\nu \approx-1$ (bottom panel) taken at $B=-20$~mT. 
    ({\bf d}) Schematic of the spin-selective attractive interaction between an optically-generated exciton in the $K^+$ valley and a $K^-$ valley hole.
    ({\bf e}) Circular-polarization-resolved reflectance spectrum at $\nu=-1$ and $D=0$. The AP resonance is fully polarized, indicative of full hole spin-valley polarization in the system.
    ({\bf f}) Magnetic hysteresis loops with (black) and without (red/blue) continuous-wave, circularly-polarized excitation at the attractive polaron (AP) resonance. A 3~$\mu$W pump at 1124~meV drives optical spin orientation in the pumped valley. 
    ({\bf g}) Steady-state spin polarization following 1124~meV excitation pulses of variable duration and power, controlled via an acousto-optic modulator (AOM) and a variable optical attenuator (VOA). Before each measurement, spins are optically initialized to the spin-up state using strong CW excitation.
    \label{fig:Fig1}}
\end{figure*}

We carried out our experiments in a gate-tunable t-MoTe$_2$ device (Figs.~\ref{fig:Fig1}{\bf a,b}) with a twist angle of around $\theta_t \simeq  3.5^\circ$. It is well established that t-MoTe$_2$ with $3.0 \le \theta_t \le 4.1$ exhibits a honeycomb moir\'e lattice with flat bands carrying valley-contrasting Chern numbers $C=\mp1$~\cite{Anderson_Science_2023,Cai_Xu_2023,Zeng_Nature_2023,Xu_PRX_2023,Park_Xu_2023,Wang_PRL_2024}. In our structure, the chemical potential $V_{\mu}$ that determines the hole filling factor $\nu$, and the displacement field $V_E$ are independently controlled using voltages applied to top ($V_T$) and bottom ($V_B$) gates (Fig.~\ref{fig:Fig1}{\bf a}). Unless indicated, all measurements are carried out in a dilution refrigerator with free-space optical access and an electron temperature estimated to be in the range $\sim0.1$--$0.2$~K (see Methods Sec.~2 for details). 

In t-MoTe$_2$, spontaneous time-reversal symmetry breaking and the associated complete valley polarization of holes for $0.4 \le \nu \le 1.3$ leads to a stark asymmetry in the optical response to right- ($\sigma^+$) and left-handed ($\sigma^-$) circularly polarized weak resonant probes. Specifically, when the holes are initialized in the $K^-$ valley by applying a small magnetic field ($B= -20$~mT), the $\sigma^+$ reflection spectrum exhibits two dominant attractive polaron (AP) resonances upon hole doping, whereas the $\sigma^-$ spectrum is dominated by weak layer-tagged exciton resonances ($X^*$) (Figs.~\ref{fig:Fig1}{\bf c,e}). By tracking these $\sigma^+$ polarized AP resonances as a function of the absolute hole density $n$ (calculated via a parallel-plate capacitor model, see Ref.~\cite{Huber_Nature_2026} for details), we can precisely determine the moir'e filling factor $\nu$. The prominent cusps in this optical response (Fig.~\ref{fig:Fig1}{\bf c}) indicates the incompressibility of the fractional (FCI) and integer (ICI) Chern insulator states. This allows us to determine the hole density corresponding to $\nu=-2/3$ and $\nu=-1$, which are $n=(-2.12 \pm 0.13)10^{12} \textrm{cm}^{-2}$ and $n=(-3.18 \pm 0.13)10^{12} \textrm{cm}^{-2}$, respectively.

The strict selection rule results in perfect $\sigma^+$ ($\sigma^-$) polarization $\rho_{\textrm{AP}}=1$ ($\rho_{\textrm{AP}}=-1$) of the AP resonance if all holes in the optically detected spot are completely spin-valley polarized in the $K^-$ ($K^+$) valley (Fig.~\ref{fig:Fig1}{\bf d,e}). Consequently, $\rho_{\textrm{AP}}$ is a direct proxy for the system's degree of spin-valley polarization, $\langle S_z \rangle$ \cite{Huber_Nature_2026}. Previous experiments have demonstrated that, upon increasing the power of incident $\sigma^+$ ($\sigma^-$) polarized light tuned to the AP (X$^*$) resonances, the holes that are initially in the $K^-$ valley are optically pumped to the $K^+$ valley, which in turn leads to vanishing AP (X$^*$) response in $\sigma^+$ ($\sigma^-$) polarization; that is, the holes are optically pumped into the dark-state of driven-dissipative evolution \cite{Huber_Nature_2026, Holtzmann_Nature_2026,cai_optical_2026}. 

Ferromagnetic hysteresis in magnetic field scans can be optically probed using low-power ($\approx 2$~nW) white light excitation of the AP resonances (Fig.~\ref{fig:Fig1}\textbf{f}). In contrast to this, pumping the AP resonance at 2~$\mu$W for a few seconds and then probing the system with low-power white light leads to a collapse of the magnetic bistability. Below the coercive field $B_c$, hole spin orientation becomes entirely dictated by the pump polarization, allowing deterministic spin reversal. We emphasize that the optically dictated spin orientation is stable for at least up to 6 hours after the preparation, provided $|B| \ll B_c$~\cite{Huber_Nature_2026}. 
In contrast, for $|B| \ge B_c$, optical orientation using laser pulses is ineffective in determining the spin polarization in steady-state. The gradual change in hole spin polarization as the magnetic field is tuned across $B_c$ at $T \simeq 0.2$~K suggests pinning of domains due to disorder.

Fig.~\ref{fig:Fig1}\textbf{g} presents the steady-state spin-valley polarization at $\nu = -1$ following optical pumping of the AP resonance at 1124~meV with a $\sigma^-$ polarized pulse of variable power and duration. Prior to each measurement, the system is initialized in the $\langle S_z \rangle = +1$ state via a strong $\sigma^+$ polarized pump pulse. The excitation pulse duration and power are independently controlled using an acousto-optic modulator (AOM) and a variable optical attenuator (VOA) (see Ext. Data Fig.~\ref{fig:SI_Fig1}). Interestingly, for a fixed pulse energy, optical spin orientation is more efficient with shorter pulses, revealing a nonlinear dependence on peak pump power.\\

{\bf Optically generated bubble collapse and expansion}

In Ising ferromagnets, the decay of a false vacuum (metastable) state  takes place by spontaneous nucleation, proliferation, and coalescence of true vacuum (stable) domains. For sample temperatures well below the exchange energy J ($k_BT \ll J$), and for $B \ll B_c$, it can be argued that the corresponding decay time is likely to be dominated by the spontaneous formation of true vacuum bubbles, whose radii exceed the critical radius. To investigate the dynamics of a single bubble with an externally controlled radius, we adopted a scheme where we optically generate a sub-wavelength region in the true vacuum state. By exploiting the nonlinear saturation of the spin-switching threshold within the spatially varying profile of a LG beam, we achieve sub-diffraction spatial confinement conceptually analogous to stimulated emission depletion (STED) microscopy \cite{Hell_OptLett_1994}. We describe the resulting configuration as a true vacuum bubble deterministically injected into the false vacuum background.

\begin{figure*}[t]
	\includegraphics[width=0.99\textwidth]{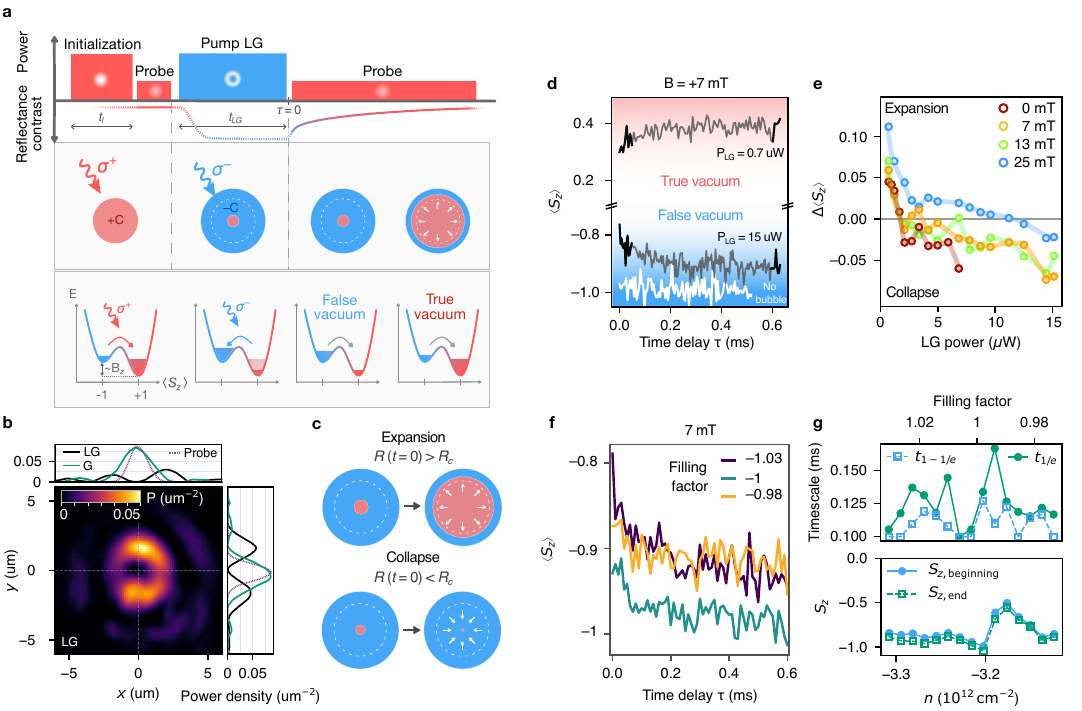}
    \caption{{\bf Collapse and expansion of an optically-generated true vacuum bubble.}
    ({\bf a}) Schematic of the pump-probe sequence. The system is initialized into a spin-up true vacuum state via a Gaussian pump pulse (10~ms, 1.5~$\mu$W). After a weak Gaussian probe (300~$\mu$s, $<2$~nW) establishes the baseline reflectivity, an oppositely polarized Laguerre-Gauss (LG) pulse with variable power and duration, drives the surrounding region into false vacuum. A final weak Gaussian (detection area indicated by the white dotted circle) monitors the subsequent evolution.
    ({\bf b}) CCD image of the reflected LG beam, alongside $x$- and $y$-axis cross-sectional profiles of the LG pump, Gaussian pump, and effective probe spots.
    ({\bf c}) Schematic of an optically injected true vacuum bubble collapsing or expanding, depending on whether its initial size is below or above the critical radius.
    ({\bf d}) Time-resolved spin polarization $\langle S_z(\tau) \rangle$ at $B = 7$~mT for low (0.7~$\mu$W) and high (15~$\mu$W) LG powers. A control trace (white) pumping with co-polarized Gaussian and LG pulses pumping into the false vacuum ($\langle S_z\rangle = -1$) confirms the signal originates exclusively from the central bubble.
    ({\bf e}) Field and power dependence of the dynamics, quantified by the net polarization change $\Delta \langle S_z \rangle$ over the first 0.6~ms (difference between late- and early-time averages in \textbf{d}). Positive (negative) values denote collapse (expansion). Higher fields require greater LG power to induce collapse, reflecting a reduced critical radius.
    ({\bf f}) Time traces at 7~mT ($B\ll B_c$) showing bubble collapse for filling factors $\nu=-0.98,-1,-1.03$.
    ({\bf g}) Characteristic decay timescales (top) and initial/final spin polarizations (bottom) extracted from (\textbf{f}).}
    \label{fig:Fig2}
\end{figure*}

Figure~\ref{fig:Fig2}\textbf{a} outlines the pump-probe scheme we developed. First, a $\sigma^+$ polarized Gaussian beam prepares the hole spins in the $K^+$ valley; unless we state otherwise, the generated state is in true vacuum with complete valley polarization. Photons scattered from a weak $\sigma^+$ polarized probe Gaussian pulse and detected using a superconducting single photon detector (SSPD) allow us to determine the reflectivity around the AP resonance that corresponds to complete $K^+$ valley polarization. We then send a $\sigma^-$ polarized LG pump pulse possessing a broader spatial extent than the probe, that switches the surrounding holes into the $K^-$ valley. Crucially, the optical vortex at the center of the LG beam leaves a localized region in the true vacuum. The spatial extent of this central bubble can be controlled by the LG pulse intensity and reduced to substantially sub-wavelength dimensions. Subsequently probing the system with a weak, non-perturbative $\sigma^+$ polarized Gaussian probe beam then allows us to determine the expansion or collapse of this optically generated bubble (see Methods Secs.~7-8 and Ext. Data Fig.~\ref{fig:SI_Fig4} for details). Figure~\ref{fig:Fig2}\textbf{b} shows the beam profile of the LG laser pulse, which exhibits a maximum azimuthal intensity variation of a factor of 2.5. However, because the LG pump is operated deep in the nonlinear saturation regime -- well above the spin-switching threshold everywhere except precisely at the vortex core -- this azimuthal asymmetry is heavily suppressed in the final spin distribution, yielding a central true vacuum bubble that is only slightly elliptical (see Methods Sec.~8 and Ext. Data Figs.~\ref{fig:fig_lg}). The actual size of the bubble depends not only on the intensity of the LG pulse but also on the diffusion of APs.

We study the evolution of the optically generated true vacuum bubble as a function of LG laser power for $\nu=-1$ using the pump-probe scheme outlined in Fig.~\ref{fig:Fig2}\textbf{a}. Figure~\ref{fig:Fig2}\textbf{c,d} describes two complementary experiments carried out at $|B| = 7$~mT $\ll B_C$: when we set the LG pulse power to $P_{LG} = 15~\mu$W, we observe that within $0.2$~ms after the LG pulse, the bubble radius decreases due to surface tension, thus increasing the size of the false vacuum region. We conclude that the radius $R$ of the optically generated bubble in the true vacuum state for this $P_{LG}$ is smaller than the critical bubble radius $R_c$. In stark contrast, for $P_{LG} = 0.7 \mu$W, we observe that the bubble expands, indicating that $R > R_c$. In both cases, the fast dynamics depicted in Fig.~\ref{fig:Fig2}\textbf{d} saturates before the bubble completely collapses or expands. We attribute this observation to disorder: if the surface tension of the domain wall is insufficient to overcome the pinning barrier, further expansion or collapse of the bubble will no longer be possible. We therefore interpret this rapid dynamics as motion of the domain wall through clean regions of the sample.

To ensure that the observed partial collapse or expansion originates from the optically injected bubble, we carried out a test experiment where we monitored $\langle S_z (\tau) \rangle$ after preparing the false vacuum state using a combination of co-circularly polarized Gaussian and LG pulses. The corresponding data in Fig.~\ref{fig:Fig2}\textbf{d}, labeled as {\sl no bubble}, shows no deviation from the initial perfect spin polarization, indicating the total absence of false vacuum decay. This measurement clearly shows that neither the motion of the outer boundary of the false vacuum, nor the spontaneous formation of bubbles or any true vacuum remnants within the pumped spot, leads to measurable changes in $\langle S_z \rangle$ for $|B| \ll B_c$.

Since $R_c$ is determined by an interplay between the surface tension that favors collapse and the bulk energy that favors expansion of the true vacuum, increasing $B$ should result in an increase in $P_{LG}$ necessary to cross-over from an expanding to a collapsing bubble. Figure~\ref{fig:Fig2}\textbf{e} shows the change in spin polarization $\Delta S_z$ during the first $0.6$~ms following the LG pulse for four different $B$ values: as predicted, increasing $B$ results in an increase in the threshold $P_{LG}$ above which the optically generated true vacuum bubble is forced to collapse. 

A natural question that the measurements in Fig.~\ref{fig:Fig2}\textbf{d,e} raise is the dependence of the bubble collapse on the filling of the electronic state. Figure~\ref{fig:Fig2}\textbf{f} shows the collapse dynamics for $\nu=-1.03, -0.98$ along with $\nu=-1$: we find that the extracted decay time does not depend strongly on $\nu$ (Fig.~\ref{fig:Fig2}\textbf{g} top panel). The main dependence on $\nu$ appears to come from the magnitude of $\langle S_z (\tau=0) \rangle$ immediately after the LG  pulse is turned off (Fig.~\ref{fig:Fig2}\textbf{g} bottom panel), suggesting that the size of the center bubble is smallest at $\nu=-1$ ---possibly due to more efficient optical pumping~\cite{Huber_Nature_2026}. \\ 

{\bf $B$ and $\nu$ dependence of bubble dynamics}

The Chern bands in t-MoTe$_2$ exhibit robust Ising ferromagnetism over a wide range of $\nu$, encompassing both compressible metallic and incompressible FCI and ICI states. Recent theoretical work has highlighted the contribution of chiral edge modes in Chern insulator states to surface tension, suggesting that the bubble dynamics could be influenced by the underlying electronic state~\cite{Fabian_Arxiv_2026}. At $|B| \ll B_c$, however, we observe no pronounced dependence of the dynamics on $\nu$ (Fig.~\ref{fig:Fig2}\textbf{f}). Instead, the limited expansion and collapse of the true-vacuum bubble indicate that domain-wall pinning of the true vacuum bubble by disorder plays an important role in this regime (Fig.~\ref{fig:Fig2}\textbf{d}). In contrast, the measurements that we describe in this section reveal distinct dynamical regimes associated with bubble nucleation and domain-wall depinning, with their relative importance depending on the magnetic field and the electronic state.

\begin{figure*}[t]
	\includegraphics[width=0.99\textwidth]{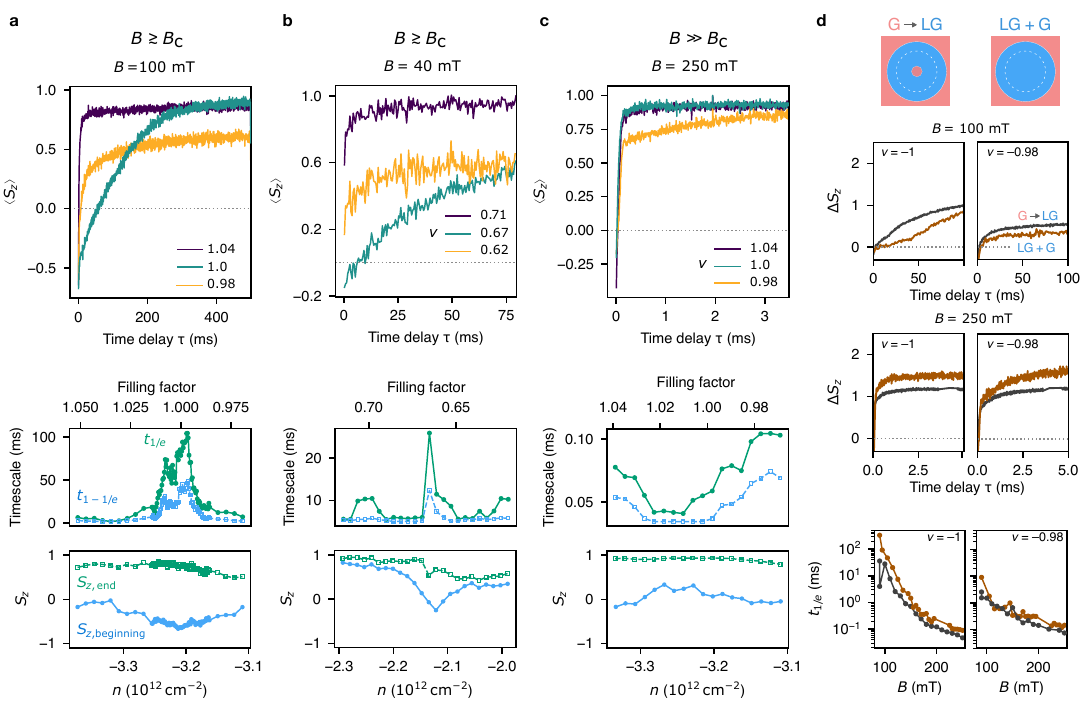}
    \caption{\textbf{Filling and magnetic-field dependence of false vacuum decay above the coercive field.} 
    ({\bf a-c}) Seeded dynamics following \(G_{TV}\!\rightarrow LG_{FV}\) preparation: a \(\sim1~\mu\mathrm{W}\), \(1\) ms Gaussian pulse initializes the true vacuum, followed by a \(15~\mu\mathrm{W}\), \(10~\mu\mathrm{s}\) LG pulse preparing the surrounding false vacuum. Top panels show representative polarization traces. The center  bottom panels show the characteristic times \(t_{1-1/e}\) and \(t_{1/e}\) where $1-1/e$ and $1/e$, respectively, of the polarization change has occurred. Bottom panels show the polarization immediately after preparation and at the end of the dynamics, respectively. Pronounced lifetime maxima occur near the integer Chern-insulating state at 100 mT (\textbf{a}) and the fractional state near $|\nu|=2/3$ at 40 mT (\textbf{b}). Far above the coercive field at 250 mT (\textbf{c}), the decay is sub-millisecond and the integer-filling lifetime enhancement disappears. 
    ({\bf d}) Comparison of preparation with a central true vacuum seed ($G_{TV} \rightarrow LG_{FV}$, gray) and without a central seed ($G_{FV}+LG_{FV}$, brown), illustrated schematically above. Representative traces at 100 and 250 mT are shown for $\nu=-1$ and $-0.98$ and corresponding decay timescales $t_{1/e}$ are plotted against field below. The seed accelerates the decay in the intermediate-field regime at $\nu = -1$ and to a lesser degree at $\nu = -0.98$. At large fields, seeded and unseeded decays approach similar time scales.
    $\Delta S_z$ corresponds to the change in spin polarization after the first recorded time delay bin. An eventual $\Delta S_z = 2$ therefore implies a complete reversal from $\langle S_z \rangle= -1$ to $+1$. $\Delta S_z < 2$ therefore occurs when either $|S_z^{\mathrm{end/beginning}}|<1$ or when conversion occurs faster than the time resolution of the measurement ($\sim 1$ $\mathrm{\mu}$s, depending on the recorded time delay window). Time traces of seeded and unseeded dynamics at several other fields are presented in Ext. Data Fig. \ref{fig:SI_seeded_vs_unseeded}.
    \label{fig:Fig3}}
\end{figure*}

At $B \gtrsim B_c$, we probe the dynamics of $\langle S_z \rangle$ over substantially longer timescales. Figure~\ref{fig:Fig3}\textbf{a} (top) shows the polarization decay near $\nu=-1$ at $B=100$mT, slightly above the steady-state $B_c$ of the ICI state at $\nu=-1$. The decay is strongly suppressed at commensurate filling of the moiré lattice (Fig.\ref{fig:Fig3}\textbf{a}, middle): the conversion timescale at $\nu=-1$ is approximately one and two orders of magnitude longer than at $\nu=-0.98$ and $\nu=-1.04$, respectively. Since the initial polarization $\langle S_z(\tau=0) \rangle$ remains comparable across this range of fillings, (Fig.~\ref{fig:Fig3}\textbf{b}, bottom panel), the prolonged decay at $\nu=-1$ is unlikely to result from substantial differences in the initial size of the optically generated true-vacuum bubble. 

To understand this striking observation, we first address whether it is indeed the proximity to commensurate filling that leads to prolonged decay times. Figure~\ref{fig:Fig3}\textbf{b} (top panel) shows $\langle S_z (t) \rangle$ for $\nu$ values around the topologically-ordered FCI state ($\nu=-2/3$) at $B= 40$~mT, which is slightly higher than $B_c$ at $\nu=-2/3$ (see Ext. Data Fig. \ref{fig:SI_Fig3}). Similar to the ICI case, we observe a sharp increase in the $\langle S_z \rangle$ decay time when the hole density approaches $\nu = -2/3$. The size of the optically prepared false vacuum region decreases as the filling factor approaches commensurate filling (consistent with more efficient optical pumping at $\nu = -2/3$ \cite{Huber_Nature_2026}): we note however, that the peak in the relaxation timescale neither coincides with the smallest initial bubble size nor appears to be uniquely determined by the latter, suggesting anomalous decay dynamics in addition to differences in optical preparation efficiency. Crucially, we exploit this extreme sensitivity of the reversal dynamics to recalibrate the charge densities corresponding to the integer and fractional filling factors, achieving an accuracy that far exceeds standard steady-state optical spectroscopy.

For $B \gg B_c$, we expect domain-wall pinning to be relatively unimportant and the dynamics is likely to be determined primarily by the speed of field-driven domain-wall expansion or contraction, which may depend on the compressibility~\cite{Fabian_Arxiv_2026}. Figure~\ref{fig:Fig3}\textbf{c} (top panel) shows $\langle S_z (\tau) \rangle$ for $\nu=-1.03, -0.98$ along with $\nu=-1$ for $B = 250$~mT: we find that the measured decay time (Fig.\ref{fig:Fig3}\textbf{c} middle panel) does not  change appreciably when $\nu$ deviates from the precise commensurate filling corresponding to ICI state.  This is consistent with our finding in Fig.~\ref{fig:Fig2}\textbf{f,g} and indicates that domain wall speed has a weak dependence on $\nu$.

With the exception of the test data in Fig.~\ref{fig:Fig2}\textbf{d} (labeled as {\sl no bubble}), all of the experiments we presented measured false vacuum with a center seed of true vacuum. To explicitly isolate the role of the optically injected true vacuum bubble for $B \ge B_c$, we compared this seeded dynamics (prepared via cross-polarized G $\rightarrow$ LG pulses) against unseeded false vacuum decay (prepared via co-polarized G + LG pulses). Figure~\ref{fig:Fig3}\textbf{d} shows the false vacuum dynamics of these two pumping configurations for $\nu = -1$ and $\nu=-0.98$. At a large field of $B=250$~mT, the seeded and unseeded decay timescales behave very similarly, indicating that the absence of an initial bubble is not a limiting factor for the false vacuum decay rate. At an intermediate field of $B=100$~mT, however, the seed strongly accelerates the false vacuum decay. For the unseeded ICI state at $\nu=-1$, we observe a prolonged, plateau-like metastability for $\tau \lesssim 50$~ms, reflecting the substantial thermodynamic bottleneck for spontaneous bulk nucleation. Crucially, the contrast in decay dynamics between the seeded and unseeded decay is reduced at $\nu=-0.98$, indicating that spontaneous bubble nucleation occurs much more readily upon doping away from the commensurate integer filling. \\

{\bf Conclusion and Outlook}

Our nonlinear optical pump-probe protocol enables the deterministic injection of a sub-wavelength true vacuum bubble within a metastable false vacuum background in t-MoTe$_2$. At fields far below the critical field, we can control the expansion or collapse of the true vacuum bubble by optically tuning its initial size and the magnetic field favoring its expansion. At fields comparable to the steady-state coercive field, the dynamics become strongly dependent on filling, with substantially prolonged decay times at the commensurate fillings of the ICI and FCI states. Comparing optically seeded and unseeded false-vacuum regions further reveals a filling-dependent contribution of bulk nucleation to the decay in this regime. By contrast, at fields well above or below the coercive field, we observe only a weak dependence of the magnetic dynamics on filling.

The different behaviors seen at low, near-coercive, and high fields can be rationalized in terms of distinct regimes of metastable dynamics. For fields far below and far above the coercive field, the dynamics probed is likely that of individual domain wall motion. At fields $B\ll B_c$, the fast dynamics saturates on timescales of order $0.2 ~\rm ms$, likely due to pinning from disorder. We interpret this rapid dynamics as stemming from the motion of the domain wall through `clean' regions of the sample with length-scales of tens of moiré lattice constants.  A potential explanation of the weak filling-dependence of timescales for this dynamics is the presence of chiral edge states at the domain wall: although the two-dimensional bulk is gapped at commensurate filling, the wall hosts two co-propagating modes associated with the two valleys (see Methods Sec.~9 for Hartree-Fock numerics on the domain wall structure). These modes host gapless excitations of both spin species, providing substantial low-energy phase space for intervalley scattering locally at domain walls.
For $B\gg B_c$, on the other hand,  domain wall pinning is expected to be relatively unimportant, and so the timescale controlling the dynamics is again linked to the kinematics of domain wall motion, with similar time scales for the commensurate and incommensurate fillings.

In contrast to the low- and high-field regimes, the much stronger filling dependence for $B\gtrsim B_c$ is consistent with reduced domain-wall pinning and/or enhanced bulk nucleation away from $\nu=-1$. As demonstrated in Fig.~\ref{fig:Fig3}\textbf{d}, the contrast between seeded and unseeded dynamics is smaller when doping away from $\nu=-1$, suggesting a larger contribution from spontaneous bubble formation and growth. We note that the hysteresis data also shows a modest difference, with a broader range of switching fields away from $\nu=-1$. The interpretation that doping an underlying incompressible Chern insulator state state plays a central role is also supported by the qualitatively similar tendency toward sharper hysteresis and slower near-coercive dynamics at $\nu=-2/3$. 

A candidate explanation for this observation is that the doped ferromagnets experience  distinct free-energy landscapes. Recall that even while the Hall conductance remains close to quantized, deviations from exact incompressible filling can be accommodated by puddles of doped quasiparticles localized in the disorder landscape, potentially generating a more heterogeneous local electronic response. In this scenario, while both doped and commensurate fillings experience the same underlying electrostatic potential --- the most likely source for domain wall pinning -- the doped state will experience a broader and more heterogeneous local magnetic energy landscape, which could arise from spatially varying screening that reduces effective pinning/nucleation barriers. Similarly, a reduction of surface tension due to binding of carriers to domain walls in locally compressible regions would lower nucleation and depinning barriers even with small changes in $\nu$. This would result in a broader regime of switching fields away from commensurate fillings, consistent with the hysteresis data.  This picture also suggests that as a rare example of a gate-tunable itinerant ferromagnet, t-MoTe$_2$  offers a new route to examining the influence of compressibility and quenched disorder on metastable Ising domain dynamics.

The enhancement of false vacuum decay in a narrow window around rational filling factors is especially  striking given that transport measurements in both quantum Hall and Chern insulator states exhibit a quantized Hall conductance plateau that extends for a finite range of electron or hole doping \cite{Xu_APS_2023}. In the case of $\nu=1$ integer quantum Hall effect in GaAs, it is well known that equilibrium ferromagnetism is much more sensitive to deviations from the exact filling, since doped charges enter the system as skyrmions or anti-skyrmions, sharply suppressing the overall magnetic order~\cite{Sondhi_Skyrmions_PRB_1993,Barrett_Skyrmions_PRL_1995}. In contrast, for Chern insulators in t-MoTe$_2$, the Ising ferromagnetism also extends well beyond the finite $\nu$-range over which quantized Hall plateaus are observed. Our findings suggest that extreme sensitivity to small changes in $\nu$ away from ICI or FCI states nevertheless appears in the  {\it nonequilibrium dynamics} of ferromagnetism in Chern bands. In particular, a comparison of the value of the Hall conductance at $\nu=-1$ and $\nu=-1.04$ shows that they differ by less than $5 \%$~\cite{Park_Nature_phys_2026}, whereas their false vacuum decay timescale differs by two orders of magnitude ---  features which we qualitatively understand as the influence of non-percolating compressible regions on the pinning and nucleation of domains. Indeed, as we noted above, the extreme sensitivity of the decay timescale to doping away from commensurate filling promises a new and highly sensitive means of calibrating the electron density. Our observations  thus hint that the nonequilibrium dynamics of strongly correlated electrons promise to reveal features not readily apparent in equilibrium measurements.

\subsection*{Acknowledgments}
 This work was supported by the Swiss National Science Foundation (SNSF) under Grant No. 2000-1-240035. W.L. and X.X. were supported by the US DOE BES (DE-SC0012509) and Vannevar Bush Faculty Fellowship (Award number N000142512047). K.W. and T.T. acknowledge support from the JSPS KAKENHI (Grant Numbers 21H05233 and 23H02052), the CREST (JPMJCR24A5), JST and World Premier International Research Center Initiative (WPI), MEXT, Japan. 
S.A.P. acknowledges support from UKRI Horizon Europe Guarantee Grant No. EP/Z002419/1. O.B. acknowledges support from a Leverhulme Trust International Professorship [Grant Number LIP-202-014], Merton College, and the Clarendon Fund. We thank Anish Kumar, John Chalker, Steven H. Simon, Shivaji Sondhi, and Amine Ben-Mhenni for insightful discussions.

\subsection*{Author contributions}
K.K. and O.H. performed the measurements and analyzed the experimental data with assistance from M.K. and A.I. W.L. and X.X. fabricated the samples. O.B. and S.A.P. contributed to the theoretical interpretation and performed numerical modeling with critical input from K.K., O.H., and A.I. K.W. and T.T. provided bulk hBN crystals. K.K., O.H., O.B., S.P. and A.I. wrote the manuscript, with input from all authors.

\section*{Methods}

\renewcommand{\figurename}{Extended Data Figure}
\renewcommand{\theHfigure}{M\arabic{figure}}
\setcounter{figure}{0}

\subsection*{1. Device fabrication}
Device fabrication relied on mechanical exfoliation of hBN (NIMS), graphite, and 2H MoTe$_2$(HQ Graphene) onto Si/SiO$_2$ substrates within an argon-filled glovebox, maintaining H$_2$O and O$_2$ concentrations at $<$0.1ppm. Suitable flakes were selected for their homogeneity under an optical microscope, and atomic force microscopy (AFM) was used to measure the thickness of the hBN layers. To assemble the twisted bilayers, a single MoTe$_2$ monolayer was used, relying on the tear-and-stack technique~\cite{Cao_Nature_2018_1,Cao_Nature_2018_2}. Heterostructure assembly was carried out via a conventional polycarbonate (PC) stamp-based dry-transfer process~\cite{Zomer_APL_2014} consisting of multiple stages. Initially, the bottom hBN and the few-layer graphene back gate were picked up and deposited onto a Si/SiO$_2$ substrate heated to roughly 180$^{\circ}$C. The PC polymer was subsequently dissolved by immersing this structure in chloroform for 10 minutes. Next, a thin hBN flake, a graphene contact strip, and half of the MoTe$_2$ monolayer were sequentially picked up. The remaining half of the MoTe$_2$ was rotated to the predetermined twist angle, picked up, and then deposited onto the prepared bottom hBN and back gate. After dissolving the polymer once more, the stack was finalized by transferring the top hBN dielectric and top gate graphite over the twisted MoTe$_2$. To guarantee pristine interfaces throughout the heterostructure, AFM cleaning was performed after every drop-down step to physically expel trapped gas bubbles and residual PC. Finally, electron beam lithography combined with metal evaporation was used to pattern the gold electrodes and contact pads onto the devices.

\subsection*{2. Experimental setup}
\begin{figure*}[t]
	\includegraphics[width=0.6\textwidth]{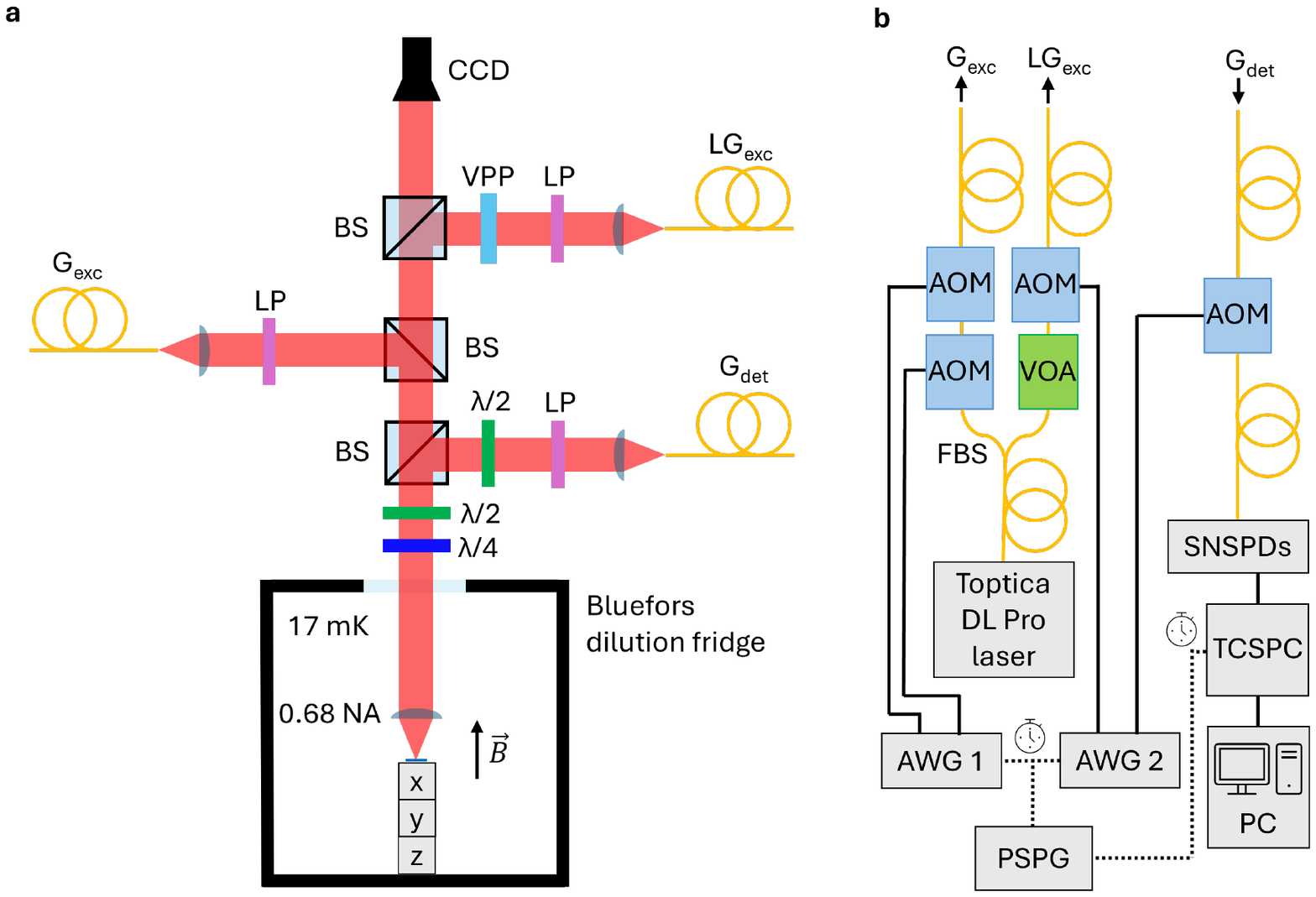}
	\caption{{\bf Schematic of the experimental setup.} 
    ({\bf a}) For the optical measurements, the light is sent from the Gaussian excitation arm and/or from the Laguerre-Gauss excitation arms to the sample. The reflected light is collected in the detection arm. All depicted lenses are aspheric. BS: beam splitter, LP: linear polarizer, $\lambda/2$: half-wave plate, $\lambda/4$: quarter-wave plate, VPP: vortex-phase plate, CCD: charge-coupled device used for imaging the sample and the beams.
    ({\bf b}) The output from a wavelength-tunable diode laser is split by a fiber-coupled beam splitter (FBS) and routed through fiber-coupled acousto-optic modulators (AOMs) to the optical setup. The signal reflected by the sample passes through an additional AOM to superconducting nanowire single-photon detectors (SNSPDs). Photon arrival times are recorded via time-correlated single-photon counting (TCSPC). A picosecond pulse generator (PSPG) acts as a master clock, providing the TCSPC timing reference and synchronizing the two arbitrary waveform generators (AWGs) that drive the AOMs.
    \label{fig:SI_Fig1}}
\end{figure*}

All measurements were performed within a cryogen-free dilution refrigerator configured for free-space optical access and integrated with a superconducting magnet capable of generating out-of-plane magnetic fields up to 9~T. The system was operated at a base temperature of approximately 17~mK, yielding effective electronic temperatures in the range of 100–200~mK~\cite{Ciorciaro_Nature_2023}. As depicted in Ext. Data Fig.~\ref{fig:SI_Fig1}\textbf{a}, incident light was focused onto the device utilizing a high-numerical-aperture aspheric lens (NA = 0.68). To ensure precise spatial targeting, the sample was mounted on $x-y-z$ piezoelectric nanopositioners providing sub-micron resolution. For the optical reflection measurements presented in Fig.~\ref{fig:Fig1}\textbf{c}, the device was excited by a broadband super-luminescent diode centered at 1155~meV. The optical head comprised two independent excitation arms (LG${\textrm{exc}}$ and G${\textrm{exc}}$) and one detection arm (G${\textrm{det}}$). Polarization optics were configured to allow the independent adjustment of the polarization state in each arm. A removable vortex phase plate (VPP) with a topological charge of $m=1$ was inserted into the LG${\textrm{exc}}$ arm to generate an LG$_{01}$ beam at the sample plane.
Ext. Data Fig.~\ref{fig:SI_Fig4}\textbf{b} presents a schematic of the experimental setup utilized for the pump-probe measurements discussed in the main text. A wavelength-tunable diode laser served as the primary light source. Its fiber-coupled output was directed through a fiber beam-splitter and subsequently routed through fiber-coupled acousto-optic modulators (AOMs) before entering the optical head. The optical signal reflected from the sample was guided through an additional fiber-coupled AOM and then delivered to a superconducting nanowire single-photon detector (SNSPD). This final AOM in the detection path was crucial to prevent saturation of the SNSPDs during excitation with high-intensity Gaussian or LG pulses. Signals from the SNSPDs were processed by a time-correlated single-photon counting (TCSPC) module and analyzed computationally. A picosecond pulse generator (PSPG) acted as the master clock for the two computer-controlled arbitrary waveform generators (AWG~1 and AWG~2) responsible for driving the AOMs. Furthermore, the PSPG supplied a phase-locked reference signal to the TCSPC module, which was a prerequisite for the temporal analysis of the pump-probe data (see Methods Sec.~6).

\subsection*{3. Determination of the optical spot size and scanner travel range}
The spatial extent of the Gaussian beam was characterized by translating the optical spot across a sharp gold electrode edge adjacent to the sample using piezoelectric actuators. The reflectance was differentiated with respect to position and fitted with a Gaussian profile, yielding a full-width at half-maximum (FWHM) of $1.5 \pm 0.1$~$\mu$m. This measurement concurrently served as a spatial calibration to map the dimensions of the CCD pixels in physical units ($\mu$m).

\subsection*{4. Image of the beams}
\begin{figure*}[t]
    \includegraphics[width=0.6\textwidth]{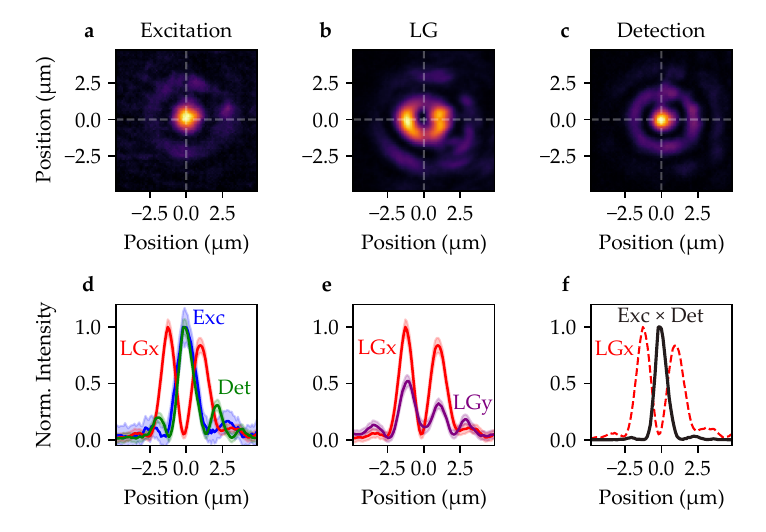}
    \caption{{\bf CCD imaging and spatial profiles of the optical beams.}
    ({\bf a}-{\bf c}) Reflected CCD images of the Gaussian excitation ({\bf a}), Laguerre-Gauss (LG) ({\bf b}), and Gaussian detection ({\bf c}) beams on the sample surface.
    ({\bf d}) Cross-sectional profiles of the three beams along the $x$-axis (horizontal axis) at $y=0$. The Gaussian excitation beam is intentionally defocused to minimize the overlap of its Airy disk with the detection beam.
    ({\bf e}) Cross-sectional profiles of the LG beam along the $x$- (LG$_x$) and $y$-axes (LG$_y$), illustrating the beam's azimuthal intensity distribution.
    ({\bf f}) Spatial product of the Gaussian excitation and detection beams along the $x$-axis, confirming negligible Airy disk contributions and minimal effective spatial overlap with the LG ring (LG$_x$).
    \label{fig:SI_Fig2}}
\end{figure*}

Prior to each measurement, global optical alignment was confirmed by imaging the reflected beams on the sample surface using a CCD camera (Ext. Data Fig.~\ref{fig:SI_Fig2}\textbf{a}-\textbf{c}). As illustrated in Ext. Data Fig.~\ref{fig:SI_Fig2}\textbf{d}, the Gaussian excitation beam is intentionally defocused to minimize the overlap of its Airy disk with that of the Gaussian detection beam. Cross-sectional profiles of the LG beam along the $x$- and $y$-axes reveal a maximum azimuthal intensity variation of approximately a factor of 2.5 (Ext. Data Fig.~\ref{fig:SI_Fig2}\textbf{e}). Finally, the spatial product of the Gaussian excitation and detection beam profiles (Ext. Data Fig.~\ref{fig:SI_Fig2}\textbf{f}) demonstrates that the Airy disk contributions are negligible, and that the overlap with the LG beam is minimal.

\subsection*{5. Coercive field dependence on the filling factor $\nu$}
\begin{figure*}[t]
    \includegraphics[width=0.6\textwidth]{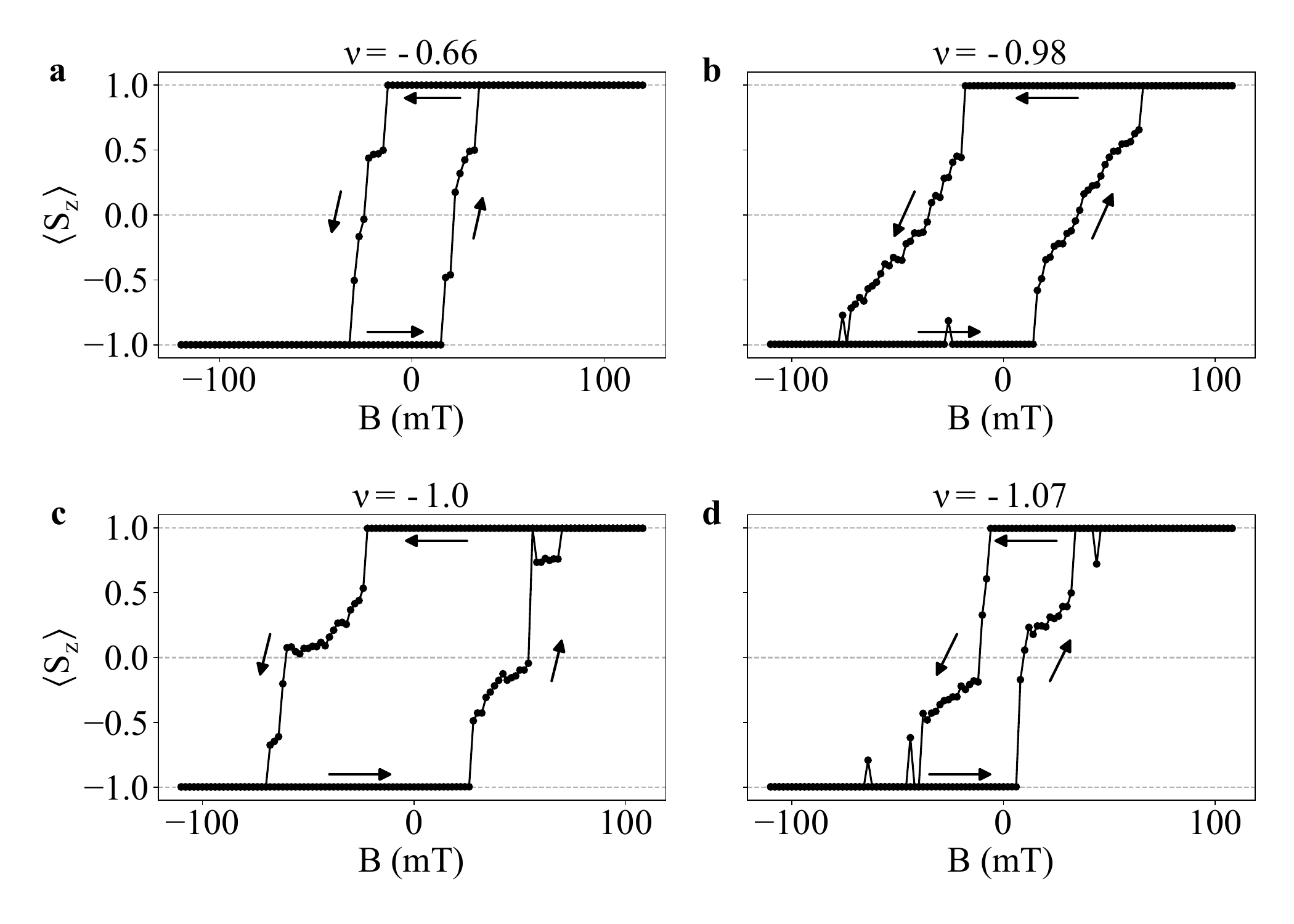}
    \caption{{\bf Hysteresis curves at various filling factors.}
    ({\bf a}-{\bf d}) Hysteresis curves at $\nu=-2/3$ ({\bf a}), $\nu=-0.98$ ({\bf b}), $\nu=-1.0$ ({\bf c}), and $\nu=-1.07$ ({\bf d}) measured with $\approx 1$~nW on the sample.
    \label{fig:SI_Fig3}}
\end{figure*}

Ext. Data Fig.~\ref{fig:SI_Fig3} displays magnetic hysteresis loops extracted from circular-polarization-resolved reflectance spectra for the fractional (FCI, $\nu=-2/3$) and integer (ICI, $\nu=-1$) Chern insulator states, alongside adjacent metallic phases at $\nu=-0.98$ and $\nu=-1.07$. The coercive field, $B_c$, exhibits a pronounced dependence on the charge density, which critically dictates the false vacuum decay dynamics observed in our time-resolved measurements. Specifically, the proximity of the applied magnetic field to $B_c$ governs the energy barrier for domain wall formation and, consequently, the rate of spontaneous bubble nucleation. For the FCI state, the reduced coercive field lowers this barrier such that an applied field of merely $B \approx 40$~mT is sufficient to drive the system into a regime where $\vert B \vert \geq B_c$  (see Fig.~\ref{fig:Fig3}\textbf{b}). In contrast, the more robust ferromagnetic order of the ICI state at $\nu=-1$ requires $\approx 80$~mT to reach this regime.

\subsection*{6. Analysis procedure of the pump-probe data}
For the pump-probe measurements detailed in Figs.~\ref{fig:Fig2} and \ref{fig:Fig3} of the main text, the experimental sequence was executed as follows (see also schematic in Fig.~\ref{fig:Fig2}\textbf{a}): First, the system was consistently initialized into the true vacuum state ($\langle S_z \rangle = +1$) via a $\sigma^+$ polarized Gaussian pulse ($\approx 1~\mu$W, 10~ms). Following this initialization, a 300~$\mu$s reference readout was performed by illuminating the sample with a weak $\sigma^+$ polarized Gaussian probe and recording the reflected optical signal. Subsequently, a $\sigma^-$ polarized LG pump pulse of variable duration and power was applied, immediately followed by the final state readout utilizing a weak $\sigma^+$ polarized Gaussian probe of variable duration. To ensure that this readout process is non-perturbative, we monitored the dynamics of a true vacuum bubble collapsing due to surface tension at 7~mT, keeping the pump parameters fixed and varying the probe power. With high probe power the bubble can be made to expand, being a bubble evolving under a continuous drive pumping into true vacuum. Upon reducing the probe power, the bubble collapses and the dynamics do not change anymore below a probe power threshold of $\lesssim 1 nW$, which constitutes the upper bound for the probe powers we used.  

The analysis procedure for the acquired datasets proceeded as follows: photons registered by the SNSPD were time-tagged and subsequently histogrammed into 1~$\mu$s temporal bins. Utilizing the synchronization signal generated by the PSPG and routed to the TCSPC module, the temporal traces from all sequence repetitions were folded onto a single period, enabling the direct accumulation of photon counts within each respective bin. Because the repetition frequency of the pump-probe cycle significantly exceeded the timescale of long-term mechanical or optical drifts within the system, the measurements were fundamentally robust against such low-frequency variations. 
The signal was subsequently normalized using a temporal window within the initial 300~$\mu$s readout block (post-Gaussian initialization). This reference window defines the baseline reflectance contrast ($\textrm{RC}=0$), representing a state where the holes are fully spin-polarized in the $K^+$ valley. To accurately compensate for count rate fluctuations induced by switching on the AOMs, the resulting time trace was further normalized against a reference trace recorded at charge neutrality ($n=0$). There, any variation in the detected photon counts originates exclusively from the instrumental response of the AOMs. To rigorously isolate this background, the reference trace was acquired employing an identical pulse and readout sequence to that of the primary measurement. Following these normalization protocols, the definitive RC is extracted as a continuous function of the probe time delay.

\subsection*{7. Quantitative extraction of spin polarization for pump-probe experiments}
The extracted RC time traces were subsequently mapped to the spin polarization $\langle S_z \rangle$. This calibration relies on the normalized reflection spectra introduced in Fig.~\ref{fig:Fig1}\textbf{c,e} of the main text. Specifically, Ext. Data Fig.~\ref{fig:SI_Fig4}\textbf{a} presents spectral linecuts of the normalized reflectance acquired at a filling factor of $\nu=-1$. From these spectra, we determine the steady-state reflectance difference between $\sigma^+$ and $\sigma^-$ polarized light at a fixed probe photon energy. This maximal reflectance contrast difference, RC$_{\textrm{max}}$, evaluated at a probe energy of 1124~meV, is plotted as a function of the filling factor $\nu$ in Ext. Data Fig.~\ref{fig:SI_Fig4}\textbf{b}. Establishing that $\textrm{RC}=0$ corresponds to the system being initialized in the true vacuum state ($\langle S_z \rangle = +1$), the temporal evolution of the spin polarization $\langle S_z \rangle (\tau)$ can be robustly quantified via the linear relation: $\langle S_z \rangle (\tau) = -2 \textrm{RC}(\tau) / \textrm{RC}_{\textrm{max}} +1$.
A representative pump-probe time trace, acquired at $\nu=-1$ and $B=100$~mT following excitation by a 15~$\mu$W, 10~$\mu$s LG pump pulse, is displayed in Ext. Data Fig.~\ref{fig:SI_Fig4}\textbf{c} (corresponding to the data shown in Fig.~\ref{fig:Fig3}\textbf{b} of the main text). This panel explicitly illustrates the dual-axis scaling, presenting both the directly measured $\textrm{RC}(\tau)$ on the primary y-axis and the derived spin polarization $\langle S_z \rangle (\tau)$ on the secondary y-axis.

\subsection*{8. Modelling Laguerre-Gauss bubble shape and size}

In this section we use a simple model, developed in Ref.~\cite{breach2026upcoming} to describe the dynamics of optical pumping \cite{Huber_Nature_2026,Holtzmann_Nature_2026,cai_optical_2026}, to qualitatively illustrate the shape of the LG seeded bubble and its dependence on power. 
We take a classical Ising model on a honeycomb lattice $H= -J \sum_{\langle i j \rangle} S_i S_j - h \sum_j h_j$, whose intrinsic (i.e. unpumped) dynamics is described by Glauber transition rates at inverse temperature $\beta$ \cite{glauber_timedependent_1963} 
\begin{eqnarray}
\label{eq:spin_flip_rate}
w_i^{\uparrow  \downarrow} = \frac{\alpha}{2} \left[ 1-\tanh{(\beta h_i)}\right] \\
w_i^{\downarrow  \uparrow} = \frac{\alpha}{2} \left[ 1+\tanh{(\beta h_i)}\right]
\end{eqnarray}
where $h_i = J \sum_{j \in nn(i)} S_i + h$ is the effective magnetic field. $\alpha$ sets the rate of the dynamics, and can be thought of as the valence band intervalley relaxation rate.

To model the pump, we introduce an excited state $X$ on each lattice site. The pump induces transitions $\uparrow~ \to X$ at a rate $W$, which is proportional to the intensity $I$ of the pump. The excited state relaxes back $X \to ~\uparrow$ at a rate $\Gamma_{\uparrow}$, and undergoes intervalley scattering before relaxation $X \to~\downarrow$ at a rate $\Gamma_{\downarrow}$. In the experimentally relevant limit where $W, \Gamma_{\downarrow} \ll \Gamma_{\uparrow}$, we can integrate out the excited state and endow the Glauber dynamics with additional spin flips at a rate $p= \frac{W\Gamma_{\downarrow}}{\Gamma_{\uparrow}} \equiv p_0 I$.

As discussed in Ref.~\cite{breach2026upcoming}, the parameters $p_0$ and $\alpha$ can be estimated by fitting the data for the pump power required to switch by $50\%$ against time (see Fig.~\ref{fig:Fig1}\textbf{g}). This yields $p_0 \sim 2 \times 10^5 \rm s^{-1} / (\mu W \mu m^{-2})$ and $\alpha \sim  4 \times 10^5 \rm s^{-1}$ (corresponding to a relaxation time of $3 \rm \mu s$, in line with previous estimations of intervalley scattering rates \cite{kim_observation_2017, cai_optical_2026}). 

With these parameters, we numerically simulate the optical pumping process with an intensity profile $I(x)$ extracted from the CCD image in Fig.~\ref{fig:SI_Fig2}\textbf{b}, taking a pulse time of $10~\rm \mu s$. In Fig.~\ref{fig:fig_lg} we show the magnetization profile for a range of total LG powers, with varying times $\tau$ after the pump is switched off. These simulations illustrate that, although the `bubble' might be extremely rough immediately after the pump is turned off, intervalley relaxation/domain-wall motion rapidly smooths the bubble into a well-defined domain for sufficiently large powers. The size of this bubble decreases as the LG power is increased, as more sites reach the `threshold' pumping level. After the well defined bubble has formed, it shrinks under curvature-driven domain wall motion. 

We emphasize that the parameters $p_0$ and $\alpha$ should be treated as rough estimates, and that the Fig.~\ref{fig:fig_lg} is a qualitative illustration rather than a quantitative prediction. Plenty of physics is missed by this simple model, including the disordered pinning landscape which we argue controls much of the observed behaviour.

\subsection*{9. Hartree-Fock Numerics for Domain Walls}

To establish the nature of the domain walls and the role of the edge states for intervalley scattering, we perform self-consistent Hartree-Fock numerics. 
At integer filling $\nu=1$, we use the standard continuum model \cite{Wu_PRL_2019,yu_fractional_2024,jia_moire_2024}, with the parameters obtained by fits to large-scale DFT in Ref.~\cite{wang_diverse_2024}, and take a double-gated screened Coulomb interaction between holes with a screening length $\xi = 25 \mathrm{nm}$ and relative permittivity $\epsilon = 20$.
To study the domain wall structure we adapt the methods of Ref.~\cite{kwan_domain_2021} to t-MoTe$_2$.  Given moiré lattice vectors $\mathbf{a}_1^M$ and $\mathbf{a}_2^M$, we assume translational symmetry in the $\mathbf{a}_2^M$ direction, and impose periodic boundary conditions but allow for translational symmetry breaking in $\mathbf{a}_1^M$. We work on a $20\times 20$ lattice, and find a self-consistent solution of the Hartree-Fock equations with two domain walls lying along the $\mathbf{a}_2^M$ direction. 

The resulting Hartree-Fock band structure (against the moiré reciprocal lattice $\mathbf{b}_2^M)$ and valley-resolved charge density are shown in Fig.~\ref{fig:fig_dw}. We find a surface tension of $\sigma=0.48~\mathrm{meV/nm}=2.8~\mathrm{meV}/a_M$, with $a_M \sim 5.8$~nm the moir\'e lattice constant.  
The domain wall is atomically thin, and does not develop intervalley coherence. This is enforced by the opposite Chern numbers of the two valleys: inter-Chern coherence requires vortices in the coherence order parameter, whose cost outweighs any exchange benefit \cite{bultinck_mechanism_2020,kwan_domain_2021,wang_diverse_2024}. 
A further consequence of the opposite Chern numbers is that the wall binds two co-propagating chiral edge states.
The sharpness and lack of intervalley coherence suggest that, under the drives of curvature or magnetic fields, the wall will move via discrete spin flips, rather than via precessional dynamics with Gilbert damping described by the Landau-Lifshitz-Gilbert equation \cite{Gilbert2004}. 
The resulting gaplessness in both valleys at the wall dramatically enhances the phase space available for intervalley scattering in comparison to a trivial insulator, which could explain why the timescales for the dynamics of commensurate $\nu=1$ and adjacent doped phases are comparable.   
This is because under a small chemical potential imbalance (due to, for example, the magnetic field), intervalley scattering can occur at the 1d Fermi surface between the two edge states as a direct, first order process.  
The electronic configuration will then relax via intravalley processes, likely dissipating excess energy via phonons. Although dissipation is expected to be much faster in the metal due to the bulk Fermi sea, if the limiting factor is the phase space for direct intervalley scattering between the Fermi surfaces of the two valleys, then there should be no significant separation in timescales for the metal and the insulator.

We also note that the edge states could lead to a sensitive dependence of surface tension on filling, which would contribute to enhanced bulk nucleation rates for the metal near coercive fields. 
When $\nu=1$ the chemical potential lies in the bulk gap, but any finite density of doped electrons or holes drives the chemical potential into the bulk bands. A fraction of these excess electrons or holes can lower their energy by occupying the wall's in-gap states, reducing the surface tension and softening nucleation and depinning barriers. 

\subsection*{Use of AI tools}
During the preparation of this work, the authors utilized ChatGPT (OpenAI), Claude (Anthropic) and Google Gemini to assist across multiple stages of the research. For data analysis, these tools were used to help generate parts of Python scripts to process and analyze the pump-probe time-traces presented in Figs.~\ref{fig:Fig2} and \ref{fig:Fig3}. During the conceptualization and drafting phases, AI tools were employed to facilitate efficient literature searches, to critique various physical interpretations against the existing literature and to improve readability of the manuscript. The authors independently verified all AI-assisted data analysis, reviewed and tested all generated code, critically evaluated all theoretical interpretations, and rigorously edited the text to ensure physical and technical accuracy. The authors take full and complete responsibility for the final content of this publication and the integrity of the presented results.

\begin{figure*}[t]
	\includegraphics[width=0.99\textwidth]{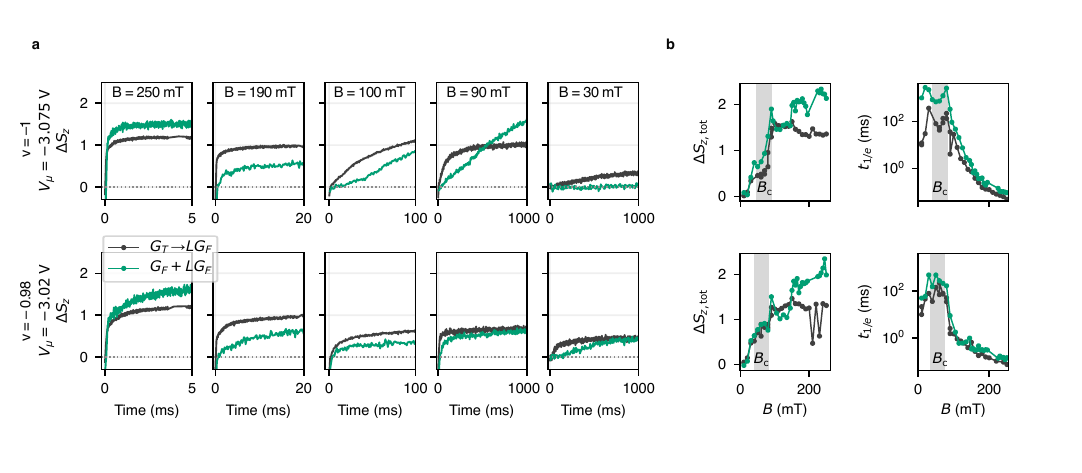}
	\caption{\textbf{Magnetic-field dependence of seeded and unseeded switching dynamics.} Top and bottom rows show measurements at \(V_\mu=-3.075\) V (\(\nu=-1\)) and \(-3.02\) V (\(\nu\simeq-0.98\)), respectively. For seeded dynamics (\(G_T\!\rightarrow LG_F\), black), a 1 ms Gaussian pulse with peak power $\sim$1 uW initializes the true vacuum before an oppositely polarized LG pulse (10 us, 15 uW) prepares the surrounding false vacuum. For dynamics without a central seed (\(G_F+LG_F\), green), both beams prepare the false vacuum (10 ms simultaneously). (\textbf{a)} Representative polarization changes as a function of time delay after the preparation at selected field magnitudes. (b)) Total measured change \(\Delta S_{z,\mathrm{tot}}\) (left) and characteristic time \(t_{1/e}\), at which \(1-1/e\) of this change is reached (right). Below the coercive field, the change in polarization $\Delta S_z^{\mathrm{tot}}$ after optical preparation drops sharply both in the seeded and unseeded case. Since the time scale shown in (\textbf{d}, bottom) refers to the timescale of the total decay $\Delta S_z^{\mathrm{tot}}$, it measures the fraction of the optical spot that still changes due to local relaxation, not the lifetime of the large area that remains metastable. Shaded regions indicate the coercive-field ranges obtained from hysteresis measurements. Note the sharp drop in $\Delta S_z$ at $\nu = -1$ for $B\sim -90$ mT in contrast to the more smooth decrease at $\nu = -0.98$, reminiscent of the steady-state spin polarization measured in hysteresis loops, see Ext. Data Fig.~\ref{fig:Fig3}. Delay windows were adapted to the strongly field-dependent dynamics.
    \label{fig:SI_seeded_vs_unseeded}}
\end{figure*}

\begin{figure*}[t]
	\includegraphics[width=0.99\textwidth]{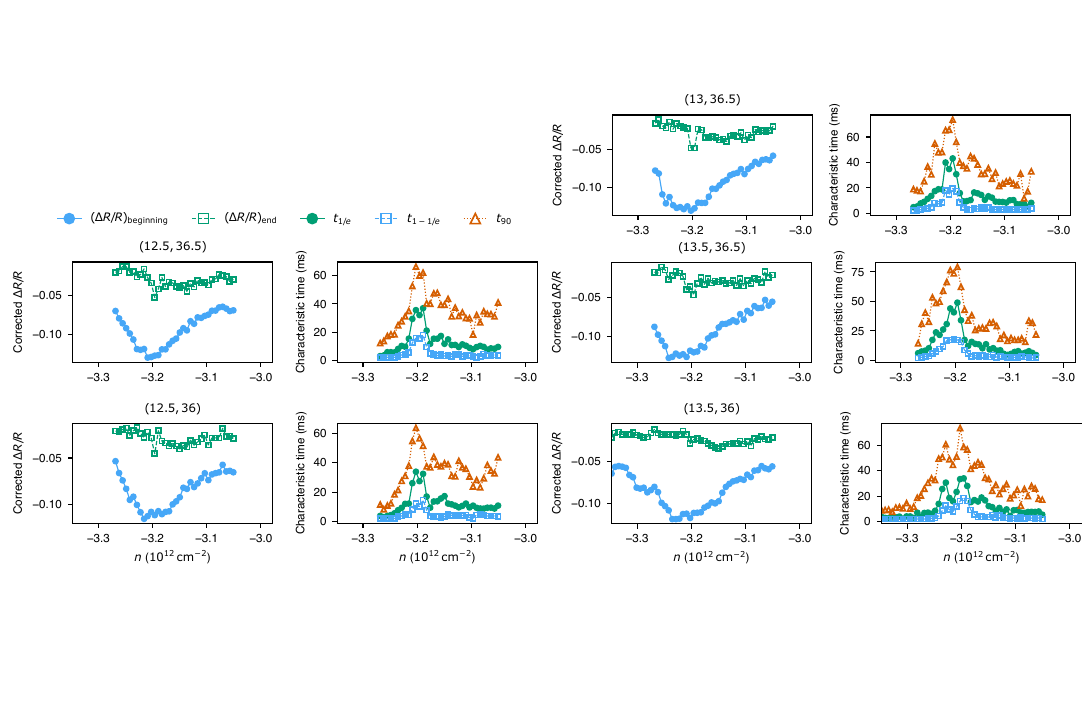}
	\caption{\textbf{Spatial reproducibility of filling-dependent switching dynamics.} Dynamics were measured at \(B=100\) mT at five nearby positions, labeled by the piezo-scanner coordinates (\(1~\mathrm{V}=0.11~\mu\mathrm{m}\), corresponding to a maximum separation of \(\sim120\) nm). The system was initialized into the true vacuum with a \(\sim1~\mu\mathrm{W}\), \(1\) ms Gaussian pulse, followed by a \(15~\mu\mathrm{W}\), \(10~\mu\mathrm{s}\) LG pulse pumping into the false vacuum. The right panel shows the characteristic percentile times \(t_{1/e}\), \(t_{1-1/e}\), and \(t_{90\%}\). The pronounced slowdown near integer filling is spatially reproducible, while the late-time dynamics exhibit position-dependent shoulders and satellite features. If the dominant contribution to the change in \(\langle S_z(\tau)\rangle\), as shown for example in Fig. \(\ref{fig:Fig3}\textbf{a}\), arises from motion of the true vacuum bubble left at the center of the LG beam, the measured contrast change and time scale are consistent with domain-wall motion across a substantial fraction of the probe. For a single sharp wall and a \(\sim 1~\mu\mathrm{m}\)-FWHM Gaussian probe, this corresponds illustratively to a displacement of several hundred nm. The spatially reproducible early and intermediate dynamics, together with the broader \(t_{90\%}\) response, are consistent with the later stages of the dynamics traversing a larger portion of the disorder and pinning landscape. These estimate is assuming a single moving domain wall; multiple domains or spatially distributed conversion could produce a similar optical signal.
    \label{fig:SI_Spatial_Variation}}
\end{figure*}

\begin{figure*}[t]
    \includegraphics[width=0.8\textwidth]{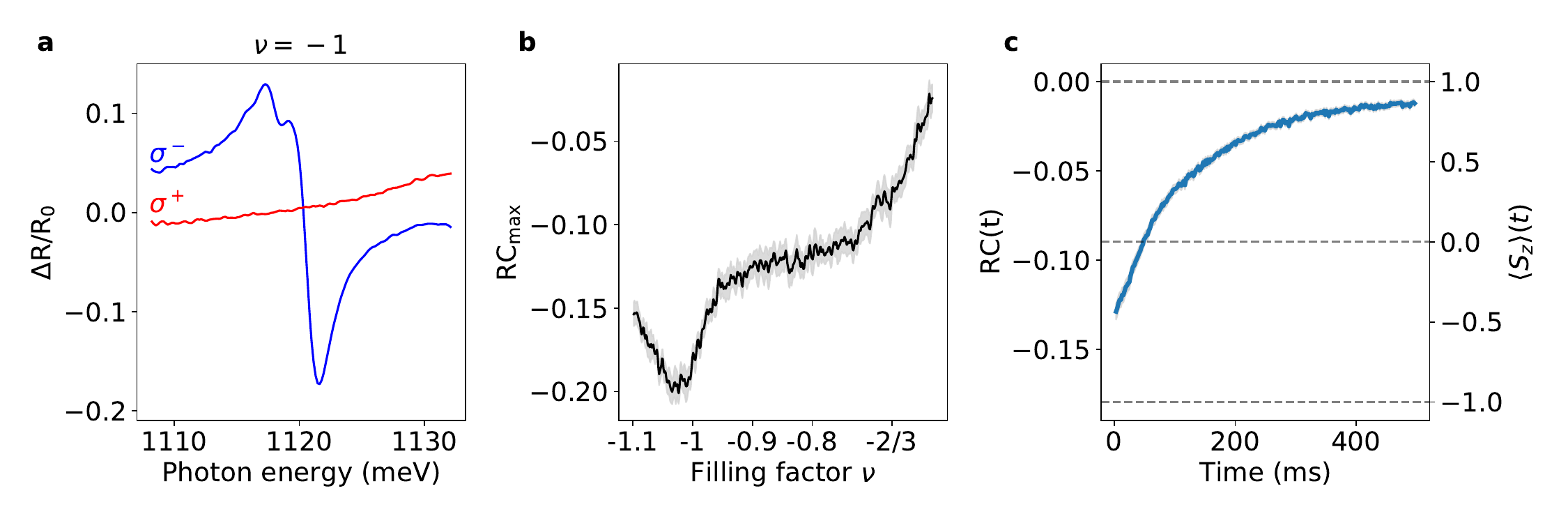}
    \caption{{\bf Extraction of $\langle S_z \rangle$ from pump-probe measurements.}
    ({\bf a}) Linecut of a smoothed, normalized reflectance spectrum at $\nu=-1$ showing the fully circularly polarized AP resonance. 
    ({\bf b}) Extracted difference in reflectance contrast between $\sigma^-$ and $\sigma^+$ polarization at a fixed photon energy (1124~meV) for filling factors $0.6 \leq \vert \nu \vert \leq 1.1$. The uncertainty on the reflectance contrast is given by the standard deviation of the raw $\Delta R/R_0$ spectra far away from the AP resonance.
    ({\bf c}) Example of a typical pump-probe time trace measured at $\nu=-1$, $B=100$~mT and after 15~$\mu$W, 10~$\mu$s LG pulses (data shown in Fig.~\ref{fig:Fig3}\textbf{b} in the main text). The RC shown on the main y-axis is obtained by normalizing with the reference counts after initialization with a Gaussian pulse. The spin polarization $\langle S_z \rangle$ shown on the secondary y-axis is then obtained using the data from panel \textbf{b}.
    \label{fig:SI_Fig4}}
\end{figure*}

\begin{figure*}[t]
    \includegraphics[width=0.7\textwidth]{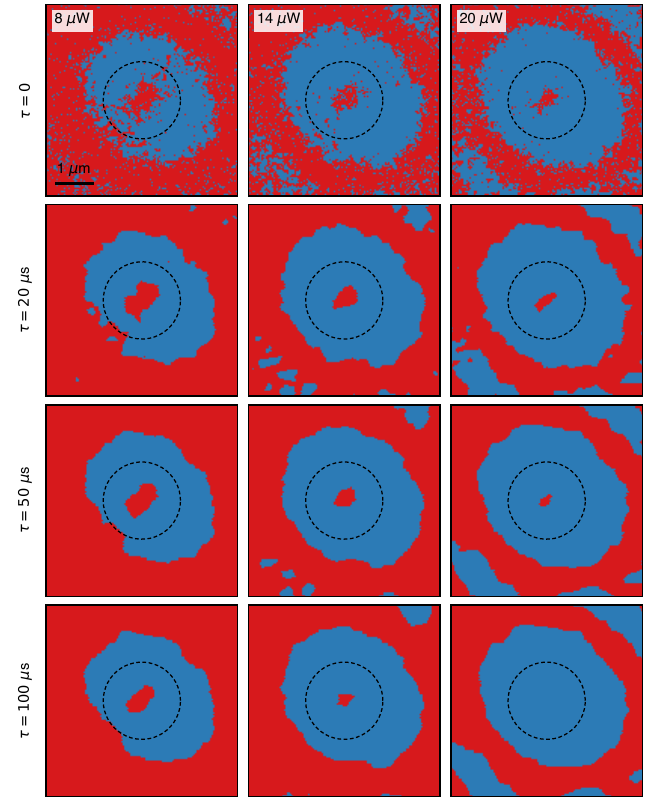}
    \caption{{\bf Illustrative magnetization profiles after Laguerre-Gauss pumping.}
    Using the simple pumped Ising model outlined in Methods Sec.~9, we simulate pumping with the measured Laguerre-Gauss intensity profile for a range of powers and a $10~\rm \mu s$ pulse, at zero applied field and temperatures $\beta J = 10$. 
    Red and blue illustrate the $K-$ (initial) and $K+$ (pumped) polarizations  respectively.
    Each column follows a single simulated evolution, with rows labelled by the time $\tau$ after the pump ends. 
    The dashed circle marks the extent of the Gaussian probe beam, which is comparable to the wavelength. 
    For sufficiently large powers, the pump produces a well-defined bubble with a size that decreases as the pump power is increased. Immediately after the pump is turned off (top row, $\tau=0$), the edge of the bubble is poorly defined, with a speckle of both polarisations. However, intervalley relaxation rapidly smooths out the boundary into a well-defined bubble on timescales of order $\tau = 20~\rm \mu s$. The bubble subsequently contracts by curvature-driven domain wall motion. At the highest powers the outer diffraction ring of the LG beam also switches, producing the detached arcs. 
    Given the simplified nature of the model, the data should be interpreted qualitatively, rather than giving a quantitative match between pumped powers and bubble shape. 
    \label{fig:fig_lg}}
\end{figure*}

\begin{figure*}[t]
    \includegraphics[width=1.0\textwidth]{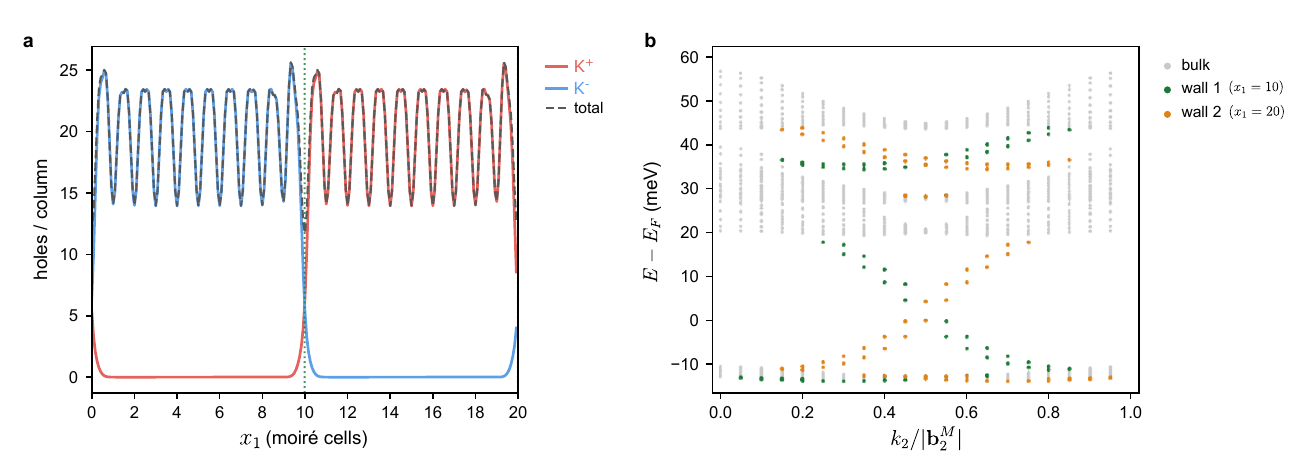}
    \caption{{\bf Domain wall structure from self-consistent Hartree-Fock.}
    In this figure we demonstrate a converged self-consistent solution to the Hartree-Fock equations with two domain walls. ({\bf{a}}) The spatially-resolved hole density as a function of position $x_1$, averaged over the $\mathbf{a}_2^M$ direction. The domain wall is atomically sharp, and does not develop intervalley coherence at the interface. Note that a small dipole spontaneously forms at the wall, breaking the reflection symmetry about the domain wall location. ({\bf{b}}) The Hartree-Fock band structure resolved by momentum parallel to the domain wall.  States with significant weight at the domain wall positions are colored in blue and green. Two copropagating chiral edge states at each domain wall cross the bulk gap.
    \label{fig:fig_dw}}
\end{figure*}

\bibliography{references}

\end{document}